\documentclass[nofootinbib,groupedaddress,superscriptaddress,unsortedaddress,showpacs]{revtex4-1}
\usepackage{amsmath,amsfonts,slashed}
\usepackage{epsfig,epsf}
\usepackage{hyperref}

\usepackage[dvips]{color}
\usepackage[normalem]{ulem}
\definecolor{darkgreen}{cmyk}{1,0,1,0.4}

\definecolor{darkred}{cmyk}{0,1,1,0.4}

\usepackage{slashed}

\newcommand{\newc}{\newcommand}
\def\preprint{{preprint}}
\def\Ord{\lower .7ex\hbox{$\;\stackrel{\textstyle <}{\sim}\;$}}
\def\OOrd{\lower .7ex\hbox{$\;\stackrel{\textstyle >}{\sim}\;$}}

\newc{\order}{{\cal O}}

\newc{\be}{\begin{equation}}
\newc{\ee}{\end{equation}}
\newc{\br}{\begin{eqnarray}}
\newc{\er}{\end{eqnarray}}
\newc{\ba}{\begin{array}}
\newc{\ea}{\end{array}}
\newc{\bi}{\begin{itemize}}
\newc{\ei}{\end{itemize}}
\newc{\bn}{\begin{enumerate}}
\newc{\en}{\end{enumerate}}
\newc{\bc}{\begin{center}}
\newc{\ec}{\end{center}}
\newc{\ul}{\underline}
\newc{\ra}{\rightarrow}
\newc{\lra}{\longrightarrow}
\newc{\wt}{\widetilde}
\newc{\til}{\tilde}

\newc{\wh}{\widehat}
\newc{\ti}{\times}
\newc{\Dir}{\kern -6.4pt\Big{/}}
\newc{\Dirin}{\kern -10.4pt\Big{/}\kern 4.4pt}
\newc{\DDir}{\kern -10.6pt\Big{/}}
\newc{\DGir}{\kern -6.0pt\Big{/}}
\newc{\sig}{\sigma}
\newc{\sigmalstop}{\sig_{\lstoppair}}
\newc{\Sig}{\Sigma}  
\newc{\del}{\delta}
\newc{\Del}{\Delta}
\newc{\lam}{\lambda}
\newc{\Lam}{\Lambda}
\newc{\gam}{\gamma}
\newc{\Gam}{\Gamma}
\newc{\eps}{\epsilon}
\newc{\Eps}{\Epsilon}
\newc{\kap}{\kappa}
\newc{\Kap}{\Kappa}
\newc{\modulus}[1]{\left| #1 \right|}
\newc{\eq}[1]{(\ref{eq:#1})}
\newc{\eqs}[2]{(\ref{eq:#1},\ref{eq:#2})}
\newc{\etal}{{\it et al.}\ }
\newc{\ibid}{{\it ibid}.}
\newc{\ibidem}{{\it ibidem}.}
\newc{\eg}{{\it e.g.}\ }
\newc{\ie}{{\it i.e.}\ }

\newc{\nonum}{\nonumber}
\newc{\lab}[1]{\label{eq:#1}}
\newc{\dpr}[2]{({#1}\cdot{#2})}
\newc{\lt}{\stackrel{<}}
\newc{\gt}{\stackrel{>}}
\newc{\lsimeq}{\stackrel{<}{\sim}}
\newc{\gsimeq}{\stackrel{>}{\sim}}
\def\lsim{\buildrel{\scriptscriptstyle <}\over{\scriptscriptstyle\sim}}
\def\gsim{\buildrel{\scriptscriptstyle >}\over{\scriptscriptstyle\sim}}
\def\lapp{\mathrel{\rlap{\raise.5ex\hbox{$<$}}
                    {\lower.5ex\hbox{$\sim$}}}}
\def\gapp{\mathrel{\rlap{\raise.5ex\hbox{$>$}}
                    {\lower.5ex\hbox{$\sim$}}}}
\newc{\half}{\frac{1}{2}}

\newc{\bQ}{\ol{Q}}
\newc{\dota}{\dot{\alpha }}
\newc{\dotb}{\dot{\beta }}
\newc{\dotd}{\dot{\delta }}
\newc{\nindnt}{\noindent}

\newc{\matth}{\mathsurround=0pt}
\def\ML{\ifmmode{{\mathaccent"7E M}_L}
             \else{${\mathaccent"7E M}_L$}\fi}
\def\MR{\ifmmode{{\mathaccent"7E M}_R}
             \else{${\mathaccent"7E M}_R$}\fi}

\newc{\mr}{\mathrm}

\newc{\siminf}{\mbox{$_{\sim}$ {\small {\hspace{-1.em}{$<$}}}    }}
\newc{\simsup}{\mbox{$_{\sim}$ {\small {\hspace{-1.em}{$>$}}}    }}

\newc {\Zboson}{{\mathrm Z}^{0}}
\newc{\thetaw}{\theta_W}
\newc{\mbot}{{m_b}}
\newc{\mtop}{{m_t}}
\newc{\sm}{${\cal {SM}}$}
\newc{\as}{\alpha_s}
\newc{\aem}{\alpha_{em}}

\newc{\ppbar}{\mbox{$p\ol{p}$}}
\newc{\bbbar}{\mbox{$b\ol{b}$}}
\newc{\ccbar}{\mbox{$c\ol{c}$}}
\newc{\ttbar}{\mbox{$t\ol{t}$}}
\newc{\eebar}{\mbox{$e\ol{e}$}}
\newc{\zzero}{\mbox{$Z^0$}}

\newc{\wplus}{\mbox{$W^+$}}
\newc{\wminus}{\mbox{$W^-$}}
\newc{\ellp}{\ell^+}
\newc{\ellm}{\ell^-}
\newc{\elp}{\mbox{$e^+$}}
\newc{\elm}{\mbox{$e^-$}}
\newc{\elpm}{\mbox{$e^{\pm}$}}
\newc{\qbar}     {\mbox{$\ol{q}$}}

\newc{\Ebar}{{\bar E}}
\newc{\Dbar}{{\bar D}}
\newc{\Ubar}{{\bar U}}
\newc{\susy}{{{SUSY}}}
\newc{\msusy}{{{M_{SUSY}}}}

\def\photino{\ifmmode{\mathaccent"7E \gam}\else{$\mathaccent"7E \gam$}\fi}
\def\taugluino{\ifmmode{\tau_{\mathaccent"7E g}}
             \else{$\tau_{\mathaccent"7E g}$}\fi}
\def\mphotino{\ifmmode{m_{\mathaccent"7E \gam}}
             \else{$m_{\mathaccent"7E \gam}$}\fi}
\newc{\gl}   {\mbox{$\wt{g}$}}
\newc{\mgl}  {\mbox{$m_{\gl}$}}

\def \chonep {{\wt\chi_1^+}}

\def \ch2p {{\wt\chi_2^+}}
\def \chonem {{\wt\chi_1^-}}
\def \ch2m {{\wt\chi_2^-}}

\def \chonjpm{{\wt\chi_j}^{\pm}}
\def \chonepm{{\wt\chi_1}^{\pm}}

\def \mchonepm{m_{\chonepm}}

\def \chtwopm{{\wt\chi_2}^{\pm}}
\def \mchtwopm{m_{\chtwopm}}
\newc{\dmchi}{\Delta m_{\wt\chi}}

\def \lspi{\wt\chi_i^0}

\def \lspone{\wt\chi_1^0}
\def \mlspone{m_{\lspone}}
\def \lsptwo{\wt\chi_2^0}

\def \lspthree{\wt\chi_3^0}

\def \lspfour{\wt\chi_4^0}

\newc{\sele}{\wt{\mathrm e}}
\newc{\sell}{\wt{\ell}}

\newc{\snue}     {\mbox{$ \wt{\nu_e}$}}
\newc{\smu}{\wt{\mu}}
\newc{\stau}{\wt{\tau}}
\newc {\nuL} {\wt{\nu}_L}
\newc {\nuR} {\wt{\nu}_R}
\newc {\snub} {\bar{\wt{\nu}}}
\newc {\eL} {\wt{e}_L}
\newc {\eR} {\wt{e}_R}
\def \slepl{\wt{l}_L}
\def \mslepl{m_{\slepl}}
\def \slepr{\wt{l}_R}
\def \mslepr{m_{\slepr}}
\def \stau{\wt\tau}

\def \sq{\wt{q}}

\newc{\msqot}  {\mbox{$m_(\sq_{1,2} )$}}
\newc{\sqbar}    {\mbox{$\bar{\wt{q}}$}}
\newc{\ssb}      {\mbox{$\squark\ol{\squark}$}}
\newc {\qL} {\wt{q}_L}
\newc {\qR} {\wt{q}_R}
\newc {\uL} {\wt{u}_L}
\newc {\uR} {\wt{u}_R}
\def \ul{\wt{u}_L}

\newc {\dL} {\wt{d}_L}
\newc {\dR} {\wt{d}_R}
\newc {\cL} {\wt{c}_L}
\newc {\cR} {\wt{c}_R}
\newc {\sL} {\wt{s}_L}
\newc {\sR} {\wt{s}_R}
\newc {\tL} {\wt{t}_L}
\newc {\tR} {\wt{t}_R}
\newc {\stb} {\ol{\wt{t}}_1}
\newc {\sbot} {\wt{b}_1}
\newc {\msbot} {m_{\sbot}}
\newc {\sbotb} {\ol{\wt{b}}_1}
\newc {\bL} {\wt{b}_L}
\newc {\bR} {\wt{b}_R}

\newc{\csquark}  {\mbox{$\wt{c}$}}
\newc{\csquarkl} {\mbox{$\wt{c}_L$}}
\newc{\mcsl}     {\mbox{$m(\csquarkl)$}}

\newc {\stopl}         {\wt{\mathrm{t}}_{\mathrm L}}
\newc {\stopr}         {\wt{\mathrm{t}}_{\mathrm R}}
\newc {\stoppair}      {\wt{\mathrm{t}}_{1}
\bar{\wt{\mathrm{t}}}_{1}}
\def \lstop{\wt{t}_{1}}

\def \lstoppair{\lstop\lstop^*}

\newc{\tsquark}  {\mbox{$\wt{t}$}}
\newc{\ttb}      {\mbox{$\tsquark\ol{\tsquark}$}}
\newc{\ttbone}   {\mbox{$\tsquark_1\ol{\tsquark}_1$}}

\newc{\mix}{\theta_{\wt t}}
\newc{\cost}{\cos{\theta_{\wt t}}}
\newc{\sint}{\sin{\theta_{\wt t}}}
\newc{\costloop}{\cos{\theta_{\wt t_{loop}}}}

\newc{\mixsbot}{\theta_{\wt b}}

\newc{\tb}{\tan\beta}
\newc{\cb}{\cot\beta}
\newc{\vev}[1]{{\left\langle #1\right\rangle}}

\newc{\mhalf}{m_{1/2}}
\newc{\mzero} {\mbox{$m_0$}}
\newc{\azero} {\mbox{$A_0$}}

\newc{\lb}{\lam}
\newc{\lbp}{\lam^{\prime}}
\newc{\lbpp}{\lam^{\prime\prime}}
\newc{\rpv}{{\not \!\! R_p}}
\newc{\rpvm}{{\not  R_p}}
\newc{\rp}{R_{p}}
\newc{\rpmssm}{{RPC MSSM}}
\newc{\rpvmssm}{{RPV MSSM}}

\newc{\sbyb}{S/$\sqrt B$}
\newc{\pelp}{\mbox{$e^+$}}
\newc{\pelm}{\mbox{$e^-$}}
\newc{\pelpm}{\mbox{$e^{\pm}$}}
\newc{\epem}{\mbox{$e^+e^-$}}
\newc{\lplm}{\mbox{$\ell^+\ell^-$}}

\def\Ecm{\ifmmode{E_{\mathrm{cm}}}\else{$E_{\mathrm{cm}}$}\fi}
\newc{\rts}{\sqrt{s}}
\newc{\rtshat}{\sqrt{\hat s}}
\newc{\gev}{\,GeV}
\newc{\mev}{~{\rm MeV}}
\newc{\tev}  {\mbox{$\;{\rm TeV}$}}
\newc{\gevc} {\mbox{$\;{\rm GeV}/c$}}
\newc{\gevcc}{\mbox{$\;{\rm GeV}/c^2$}}
\newc{\intlum}{\mbox{${ \int {\cal L} \; dt}$}}
\newc{\call}{{\cal L}}
\def \met  {\mbox{${E\!\!\!\!/_T}$}}

\newc{\ptmiss}{/ \hskip-7pt p_T}

\newc{\PT}{\mbox{$p_T$}}
\newc{\ET}{\mbox{$E_T$}}
\newc{\dedx}{\mbox{${\rm d}E/{\rm d}x$}}
\newc{\ifb}{\mbox{${\rm fb}^{-1}$}}
\newc{\ipb}{\mbox{${\rm pb}^{-1}$}}
\newc{\pb}{~{\rm pb}}
\newc{\fb}{~{\rm fb}}
\newc{\ycut}{y_{\mathrm{cut}}}
\newc{\chis}{\mbox{$\chi^{2}$}}

\def \jet(s){\emph{jet(s) }}

\newc{\mpl}{M_{\rm Pl}}
\newc{\mgut}{M_{GUT}}
\newc{\mw}{M_{W}}
\newc{\mweak}{M_{weak}}
\newc{\mz}{M_{Z}}

\newc{\OPALColl}   {OPAL Collaboration}
\newc{\ALEPHColl}  {ALEPH Collaboration}
\newc{\DELPHIColl} {DELPHI Collaboration}
\newc{\XLColl}     {L3 Collaboration}
\newc{\JADEColl}   {JADE Collaboration}
\newc{\CDFColl}    {CDF Collaboration}
\newc{\DXColl}     {D0 Collaboration}
\newc{\HXColl}     {H1 Collaboration}
\newc{\ZEUSColl}   {ZEUS Collaboration}
\newc{\LEPColl}    {LEP Collaboration}
\newc{\ATLASColl}  {ATLAS Collaboration}
\newc{\CMSColl}    {CMS Collaboration}
\newc{\UAColl}    {UA Collaboration}
\newc{\KAMLANDColl}{KamLAND Collaboration}
\newc{\IMBColl}    {IMB Collaboration}
\newc{\KAMIOColl}  {Kamiokande Collaboration}
\newc{\SKAMIOColl} {Super-Kamiokande Collaboration}
\newc{\SUDANTColl} {Soudan-2 Collaboration}
\newc{\MACROColl}  {MACRO Collaboration}
\newc{\GALLEXColl} {GALLEX Collaboration}
\newc{\GNOColl}    {GNO Collaboration}
\newc{\SAGEColl}  {SAGE Collaboration}
\newc{\SNOColl}  {SNO Collaboration}
\newc{\CHOOZColl}  {CHOOZ Collaboration}
\newc{\PDGColl}  {Particle Data Group Collaboration}

\def\issue(#1,#2,#3){{\bf #1}, #2 (#3)}
\def\iss(#1,#2,#3){{\bf #1} (#3) #2}
\def\iss1(#1,#2,#3){{\bf #1} (#2) #3}
\def\issuenew(#1,#2,#3,#4){{\bf #1} (#2) {\bf no.#4}, #3}
\def\ASTR(#1,#2,#3){Astropart.\ Phys. \issue(#1,#2,#3)}
\def\AJ(#1,#2,#3){Astrophysical.\ Jour. \issue(#1,#2,#3)}
\def\AJS(#1,#2,#3){Astrophys.\ J.\ Suppl. \issue(#1,#2,#3)}
\def\APP(#1,#2,#3){Acta.\ Phys.\ Pol. \issue(#1,#2,#3)}
\def\JCAP(#1,#2,#3){JCAP \iss1(#1,#2,#3)} 
\def\SC(#1,#2,#3){Science \issue(#1,#2,#3)}
\def\PRD(#1,#2,#3){Phys.\ Rev.\ D \issue(#1,#2,#3)}
\def\PR(#1,#2,#3){Phys.\ Rev.\ \issue(#1,#2,#3)} 
\def\PRC(#1,#2,#3){Phys.\ Rev.\ C \issue(#1,#2,#3)}
\def\NPB(#1,#2,#3){Nucl.\ Phys.\ B \iss1(#1,#2,#3)}
\def\NPPS(#1,#2,#3){Nucl.\ Phys.\ Proc. \ Suppl \issue(#1,#2,#3)}
\def\NJP(#1,#2,#3){New.\ J.\ Phys. \issue(#1,#2,#3)}
\def\JP(#1,#2,#3){J.\ Phys.\issue(#1,#2,#3)}
\def\JPG(#1,#2,#3){J.\ Phys.\ G \issue(#1,#2,#3)}
\def\PL(#1,#2,#3){Phys.\ Lett. \issue(#1,#2,#3)}
\def\ZP(#1,#2,#3){Z.\ Phys. \issue(#1,#2,#3)}
\def\ZPC(#1,#2,#3){Z.\ Phys.\ C  \issue(#1,#2,#3)}
\def\PREP(#1,#2,#3){Phys.\ Rep. \issue(#1,#2,#3)}
\def\PRL(#1,#2,#3){Phys.\ Rev.\ Lett. \issue(#1,#2,#3)}
\def\MPL(#1,#2,#3){Mod.\ Phys.\ Lett. \issue(#1,#2,#3)}
\def\RMP(#1,#2,#3){Rev.\ Mod.\ Phys. \issue(#1,#2,#3)}
\def\SJNP(#1,#2,#3){Sov.\ J.\ Nucl.\ Phys. \issue(#1,#2,#3)}
\def\CPC(#1,#2,#3){Comp.\ Phys.\ Comm. \issue(#1,#2,#3)}
\def\IJMPA(#1,#2,#3){Int.\ J.\ Mod. \ Phys.\ A \issue(#1,#2,#3)}
\def\MPLA(#1,#2,#3){Mod.\ Phys.\ Lett.\ A \issue(#1,#2,#3)}
\def\PTP(#1,#2,#3){Prog.\ Theor.\ Phys. \issue(#1,#2,#3)}
\def\RMP(#1,#2,#3){Rev.\ Mod.\ Phys. \issue(#1,#2,#3)}
\def\NIMA(#1,#2,#3){Nucl.\ Instrum.\ Methods \ A \issue(#1,#2,#3)}
\def\EPJC(#1,#2,#3){Eur.\ Phys.\ J.\ C \issue(#1,#2,#3)}
\def\RPP (#1,#2,#3){Rept.\ Prog.\ Phys. \issue(#1,#2,#3)}
\def\PPNP(#1,#2,#3){ Prog.\ Part.\ Nucl.\ Phys. \issue(#1,#2,#3)}
\newc{\PRDR}[3]{{Phys. Rev. D} {\bf #1}, Rapid  Communications, #2 (#3)}

\def\PLB(#1,#2,#3){Phys.\ Lett.\ B  \issue(#1,#2,#3)}
\def\JHEP(#1,#2,#3){JHEP \issue(#1,#2,#3)}
\def\PRDnew(#1,#2,#3,#4){Phys.\ Rev.\ D \issuenew(#1,#2,#3,#4)}

\def\gmin2{(g-2)_\mu}

\catcode`\@=11 

\def\vev#1{\left\langle #1\right\rangle}
\def\lsim{\mathrel{\mathpalette\@versim<}}
\def\gsim{\mathrel{\mathpalette\@versim>}}
\def\@versim#1#2{\vcenter{\offinterlineskip
    \ialign{$\m@th#1\hfil##\hfil$\crcr#2\crcr\sim\crcr } }}
\def\etal{{\em et. al.}}

\def\r2{\sqrt 2}
\def\beq{\begin{equation}}
\def\eeq{\end{equation}}
\def\beqn{\begin{eqnarray}}
\def\eeqn{\end{eqnarray}}
\def\sinW2{\sin^2\theta_W}

\def\mz2{M_{z}^2}
\def\c2b{\cos 2\beta}

\def\m#1{{\tilde m}_#1}

\def\mw#1{{\tilde m}_{\omega #1}}

\def\mz{M_Z}
\def\m0{m_0}
\def\mhalf{m_{\frac{1}{2}}}

\def\cb{\cos\beta}

\def\sec2w{sec^2\theta_W}

\def\gmin2{(g-2)_\mu}

\catcode`\@=11 

\def\vev#1{\left\langle #1\right\rangle}
\def\lsim{\mathrel{\mathpalette\@versim<}}
\def\gsim{\mathrel{\mathpalette\@versim>}}
\def\@versim#1#2{\vcenter{\offinterlineskip
    \ialign{$\m@th#1\hfil##\hfil$\crcr#2\crcr\sim\crcr } }}
\def\etal{{\em et. al.}}

\def\tb{\tilde b}

\def\tL{\tilde L}

\def \chonep{{\wt\chi_1}^{+}}
\def \chonem{{\wt\chi_1^-}}
\def \chonep2{{\wt\chi_2^+}}
\def \chonem2{{\wt\chi_2^-}}

\def \chonjpm{{\wt\chi_j}^{\pm}}
\def \chonepm{{\wt\chi_1}^{\pm}}

\def \mchonepm{m_{\chonepm}}

\def \chtwopm{{\wt\chi_2}^{\pm}}
\def \mchtwopm{m_{\chtwopm}}

\def \lstop{\wt{t}_{1}}

\def \lspi{\wt\chi_i^0}

\def \lspone{\wt\chi_1^0}
\def \mlspone{m_{\lspone}}
\def \lsptwo{\wt\chi_2^0}

\def \lspthree{\wt\chi_3^0}

\def \lspfour{\wt\chi_4^0}

\def\PL{Phys. Lett.}
\def\PRL{Phys. Rev. Lett.}

\def\PR{Phys. Rev.}

\def \lsptwo{\wt\chi_2^0}
\def \lspone{\wt\chi_1^0}
\def \chonem {{\wt\chi_1^\pm}}
\def \chargino1 {{\wt\chi_1^\pm}}
\def \chargino2 {{\wt\chi_2^\pm}}
\def \lstop{\wt{t}_{1}}
\def \ch2m {{\wt\chi_2^-}}
\def \lspi{\wt\chi_i^0}
\def \chonep {{\wt\chi_1^+}}

\begin{document}

\renewcommand*{\thefootnote}{\fnsymbol{footnote}}


\title{New Limits on Heavier Electroweakinos and their LHC Signatures}


\vspace{5mm}

\author{Amitava Datta}
\email{adatta\_ju@yahoo.co.in}
\author{Nabanita Ganguly}
\email{nabanita.rimpi@gmail.com}
\affiliation{Department of Physics, University of Calcutta,\\
92 Acharya Prafulla Chandra Road, Kolkata 700009, India}
\author{Sujoy Poddar}
\email{sujoy.phy@gmail.com}
\affiliation{Netaji Nagar Day College,170/436 N.S.C. Bose Road, Kolkata - 700092, India}  

\begin{abstract}

We investigate the heavier electroweakino sectors in several versions of the 
MSSM, which has not been explored so far in the light of the LHC  data, 
and obtain new bounds using the ATLAS Run I constraints in the $3l + \met$ 
channel. We also venture beyond the trilepton events and predict several novel 
multilepton + $\met$ signatures of these electroweakinos which may show up during LHC
Run II before the next long shutdown. These signals can potentially distinguish
between various models with nondecoupled heavier electroweakinos and the much studied ones 
with decoupled heavier electroweakinos.       

\end{abstract}

\date{\today}




\maketitle

\renewcommand*{\thefootnote}{\arabic{footnote}}
\setcounter{footnote}{0}



The null results from the searches for supersymmetry (SUSY)\cite{susyrev} during Run I of the LHC have imposed stringent bounds on the masses of the strongly 
interacting supersymmetric particles (sparticles)\cite{atlastwiki,cmstwiki}, some of which have been further strengthened by the preliminary results from the Run II at 13 TeV. This trend naturally provokes a close scrutiny of a scenario where all the strongly interacting sparticles are beyond the reach of the experiments before the next shutdown. If this indeed be the case then the prospective SUSY signals during the next few years will be governed by the electroweak (EW) sector. It is also worth recalling that this sector alone accounts for several phenomenological triumphs of SUSY like explanation of the observed dark matter (DM) relic density of the universe \cite{wmap,planck,dmrev1}, alleviation of the tension between the precisely measured value of the anomalous magnetic moment of the muon \cite{muonexp} and the SM prediction \cite{muonrev}.

In the  R-parity conserving Minimal Supersymmetric Standard Model (MSSM) without any assumption regarding the soft SUSY breaking mechanism, the fermionic sparticles in the EW sector are the charginos ($\chonjpm$, $j= 1-2$) and the 
neutralinos ($\lspi$, $i = 1 - 4$) - collectively called the electroweakinos (eweakinos). In the MSSM the masses and the compositions of these sparticles are determined by four independent parameters: the U(1) gaugino mass parameter $M_1$, the SU(2) gaugino mass parameter $M_2$, 
the higgsino mass parameter $\mu$  and tan $\beta$, the ratio of the vacuum expectation values of the two neutral Higgs bosons. Throughout this paper we take tan $\beta$ = 30 which usually gives a better agreement with the $\gmin2$ data. The indices j and i are arranged in ascending order of the masses. The stable, neutral lightest neutralino ($\lspone$) is a popular DM candidate.     
The scalar sparticles are the $L$ and $R$ type sleptons and the sneutrinos. We assume L (R)-type
sleptons of all flavours to be  mass degenerate with a common mass $\mslepl$ ($\mslepr$). Because of SU(2) symmetry the sneutrinos are mass  degenerate with L-sleptons modulo the D-term contribution. We neglect LR mixing in the slepton sector.
For simplicity we work in the decoupling regime of the Higgs sector of the MSSM with only one light, SM like Higgs boson 
\cite{djouadi}, a scenario consistent with all Higgs data collected so far\cite{philip}.  
 
During Run I  the eweakino searches were mainly based on the hadronically quiet $3l + \met $ signal \footnote{In this paper $l$ stands for $e$ and $\mu$ unless otherwise mentioned}. The null results from these searches were interpreted by the LHC collaborations in terms of several simplified models consisting of  a  minimal set of parameters required to study this signal. It was, e.g., assumed in all analyses that the $3l$ signal comes only from
$\chonepm$ - $\lsptwo$ production followed by their  leptonic decays \cite{atlas3l,cms3l} while the heavier eweakinos are decoupled. This resulted in correlated bounds on $\mchonepm$ and $\mlspone$. In contrast in \cite{mc1,mc2,arghya} the data was reinterpreted in terms of different MSSMs some of which are closely related to the above simplified models. In each case the full set of parameters belonging to the EW sector  are specified so that one can also address other important issues like the DM relic density, the correlation among the trilepton and slepton search data etc. 

In this letter we emphasize the potential signatures of the hitherto unexplored heavier eweakinos in the
upcoming LHC experiments at 13 TeV before the next shutdown. That these signals are indeed well within the reach of the ongoing experiments is indicated by the observation that the published bounds  on the lighter eweakinos masses from Run I turn out to be quite sensitive to the masses of heavier eweakinos. This we shall show below by relaxing the ad hoc assumption of strict decoupling. The rich phenomenology of the non-decoupled scenarios is further illustrated by some novel signatures like events with $4l$s, three same sign and one opposite sign ($SS3OS1$) leptons and $5l$s, all accompanied by large $\met$, which may be observed  with $ \lsim 100 \ifb$ of luminosity i.e., before the next long shutdown. Most important:  for  a compressed lighter eweakino spectrum all viable leptonic signals including the $3l$ events are due to the heavier ones. In addition in a wide variety of non-compressed models the source of m-lepton signals with
m $>$ 3 are the non-decoupled heavier eweakinos.

The constraints from the trilepton searches are also sensitive to the composition of the eweakinos.The analyses are mainly restricted to two generic scenarios \footnote{ We shall, however, briefly  comment on other models as well.}.

a)The Light Wino (LW) models: Many  analyses assume that the
$\chonepm$ and $\lsptwo$ are purely wino and nearly mass degenerate while the $\lspone$ is bino dominated \cite{atlas3l,cms3l,mc1}. These two lighter eweakinos have closely spaced masses governed by the  parameter $M_2$  while the $\lspone$ is either bino dominated  with mass controlled by the U(1) gaugino mass parameter $M_1$ or a bino-higgsino admixture ($ M_1 \lsim \mu$). The two heavier electroweakinos are higgsino like with masses approximately equal to $\mu$, where $M_2 < \mu$  

b)The Light Higgsino (LH) models: In contrast  this paper, following Ref. \cite{mc2}, mainly addresses scenarios with higgsino dominated $\chonepm$, $\lsptwo$ and $\lspthree$ while the LSP is either bino dominated or a bino-higgsino admixture.The three lighter eweakinos have  closely spaced masses governed by $\mu$  while the $\lspone$ is either bino dominated  with mass controlled by  $M_1$ or a bino-higgsino admixture ($ M_1 \lsim \mu$). The two heavier electroweakinos are wino like with masses approximately equal to $M_2$, where $M_2 > \mu$ 

We recall that the models belonging to class a) (b))  yield stronger (weaker) mass bounds for reasons explained in \cite{mc2}. In all models the trilepton rates  also depend sensitively on the hierarchy among the slepton and eweakino masses. If the sleptons are lighter (heavier) than $\chonepm$ and $\lsptwo$, the leptonic Branching Ratios (BR) of these eweakinos are typically large (small) yielding stronger (weaker) limits. The nomenclatures introduced in \cite{mc2} also indicate this hierarchy (e.g., Light Wino and light Left Slepton (LWLS) model, Light Higgsino and Heavy Slepton (LHHS) model etc. ). In the LHHS model both $L$ and $R$ type sleptons of all flavours are heavier than $\chonepm$ and $\lsptwo$.

We now derive the new limits in different models 
following the procedure of the ATLAS collaboration \cite{atlas3l}. 
The Tables 7 and 8 of \cite{atlas3l} contain the number of observed $3l + \met$ events and  
the SM backgrounds obtained from the data for a number of signal regions (see Table 4 in \cite{atlas3l}). Each signal region is characterized by a set of kinematical cuts.  
From these information  the model independent 95\% CL upper limit on any Beyond SM (BSM) events ( $N^{95}_{obs}$) for each signal region were 
computed and displayed in the same tables. Using these  upper bounds the ATLAS group obtained an exclusion contour in a simplified LWLS model 
(see Fig. 7a of \cite{atlas3l}). We validate our simulation by reproducing this exclusion contour and  proceed to obtain new constraints in 
several models with non-decoupled heavier eweakinos \footnote{An earlier example of the reliability of our simulation is presented in Fig. 7a of 
\cite{mc1}. See also \cite{arghya}, Fig. 6}.

We generate the SUSY spectrum using SUSYHIT \cite{susyhit} and simulate the signal events using
PYTHIA (v6.4) \cite{pythia} (for the details see \cite{mc1,mc2}). We use CTEQ6L \cite{cteq6l} for parton distribution functions in all our analyses. Jets are reconstructed by the anti-$k_t$ \cite{antikt} algorithm using FASTJET \cite{fastjet}
coupled with PYTHIA with $R=0.4$. The jets are required to have $P_T \geq 20$ GeV  and $|\eta|<2.5$. In all our analyses the following lepton selection criteria have been employed: 
i) All leptons (e and $\mu$) in the final state with pseudo-rapidity  $|\eta|<2.5$ and transverse momentum $P_T> 10$GeV  are selected. ii)Each lepton is required to pass the isolation cuts as defined by the ATLAS/CMS collaborations \cite{atlas3l,cms3l}. These selection cuts have been implemented in all analyses in this paper.

We obtain the most striking consequences in the LHHS model yielding the weakest bounds on the higgsino like lighter eweakinos - $\chonepm$, $\lsptwo$ and $\lspthree$ \cite{mc2}.
Naturally the possibility that the heavier eweakinos ($\chtwopm$ and $\lspfour$) also have relatively small masses is open in this case.  They are wino like with masses approximately equal to the SU(2) gaugino mass parameter $M_2$, where $M_2 > \mu$. For this class of models the common slepton mass is chosen to be 
$\mslepl = \mslepr = (\mchonepm + \mchtwopm)/2$ so that they are always lighter (heavier) than
the heavier (lighter) eweakinos. The slepton and the LSP masses are carefully chosen in all our analyses that they are consistent with the constraints from Run I direct slepton searches \cite{sleprun1}. 

If the lighter eweakino spectrum is compressed, i.e., $M_1$ $\approx \mu$, then $\chonepm$, $\lspone$,$\lsptwo$ and $\lspthree$ have large bino and higgsino components and are approximately degenerate. For all numerical computations we take $\mu=1.05~ M_1$. Consequently the $3l + \met$ or any other leptonic signal from $\chonepm$ - $\lsptwo$ (or $\lspthree$) production is not viable since the
energy release in each underlying  decay is small. On the other hand it is known for a long time that
an LSP which is a bino-higgsio admixture is attractive both from the point of view of the observed DM relic density of the universe and naturalness (\cite{arkani},\cite{baer}). The correlation between acceptable DM relic density and the $3l$ signal in the compressed scenario with decoupled $\chtwopm$ and $\lspfour$ ($M_2 \simeq 2 \mu $) can be understood from  the LHHS model discussed in \cite{mc2}. From Fig. 5 of \cite{mc2}, it follows that annihilation and co-annihilation of a bino-higgsino LSP produce acceptable DM relic density over a parameter space where the $3l$ signal is weak. We have checked that if the above sparticles are non-decoupled ($M_2 < 2 \mu$) the parameter space allowed by the WMAP and Planck data changes very little. Acceptable relic density, e.g., is obtained for the range $1.05 \mu \leq M_2 \leq 1.5 \mu$ a part of which also yields  novel LHC signals.\footnote{We  note in passing that the proposed LHC signatures of compressed scenarios with decoupled heavier eweakinos have so far been based on the monojet+ $\met$ topology or the vector boson fusion topology with forward jet tagging(\cite{guidice}-\cite{cms}). However, revealing the underlying physics with these signatures alone will indeed be impossible.}

The above tension eases out if the heavier eweakinos are not decoupled. In this case 
the  wino like $\chtwopm$ and  $\lspfour$ are pair produced with reasonably large cross-sections over a sizable portion of the parameter space. Moreover their 2-body leptonic decays via sleptons with large BRs are potential sources  of observable trilepton signals. Using the above ATLAS upper bounds on $N^{95}_{obs}$, we obtain  the first published exclusion contour in the $\mlspone$ - $\mchtwopm$ plane (Fig. \ref{fig1}). For $\mlspone \approx 80$ GeV below which $\mchonepm$ violates the LEP bound, there is a strong bound
$\mchtwopm > 610$ GeV. On the other hand for $\mlspone  \geq 170$ GeV,  there is no bound  on $\mchtwopm$. For  $\mchtwopm \lsim 300$ GeV, $\chtwopm$ and $\lspfour$  develop significant bino-higgsno component and the constraints weaken. Below $\mchtwopm \approx 250$ GeV all the eweakinos are approximately degenerate and the model cannot be constrained any further. 
For illustrating the signatures of this compressed model at LHC Run II  we have chosen the benchmark point BP1 which is presented in 
Table \ref{tab1} along with the corresponding bound on $\mchtwopm$.  

\begin{figure}[h]
\centering
\includegraphics[width=0.5\textwidth]{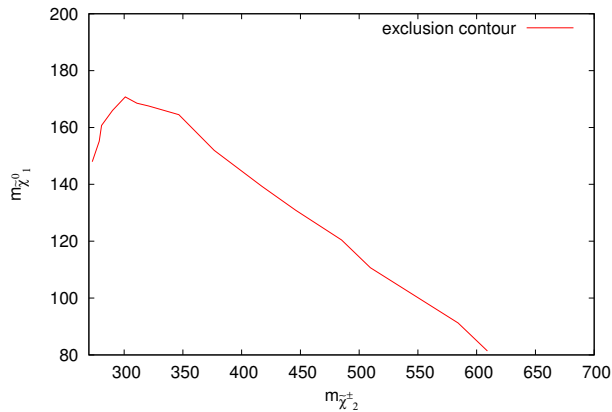}

\caption{The red contour represents the excluded parameter space in the $\mchtwopm - \mlspone$ plane using ATLAS trilepton search data from LHC RUN I. Instead of following the usual practice of considering $\chonepm - \lsptwo$ production only we have taken into account all possible eweakino pair production in the compressed LHHS model (see text for the details). All masses are in GeV.}
\label{fig1}
\end{figure}

In order to get a preliminary  estimate of the reach of Run II experiments at 13 TeV via the $3l + \met$ channel we have simulated the signal  and all  SM processes considered as backgrounds in the ATLAS $3l$ analysis reported above. The backgrounds are  suppressed by selecting events with 

\noindent
A1) 3 isolated leptons consistent with the selection cuts mentioned above,\\
A2) invariant mass of any pair of oppositely charged leptons of same flavour not in the window 81.2 $< m_{inv} <$ 101.2GeV, \\
and\\
A3) $\met >$ 200 GeV.\\ 

The total SM background is estimated to be $26.71$ for an integrated luminosity of $100 fb^{-1}$. Taking  $S/\sqrt{B} \geq 5$ to be an indicator of the observability of the signal,  we find that for $\mlspone$ = 80 (250) GeV , the reach in the compressed model is $\mchtwopm$ = 820 (672) GeV for 100 $\ifb$ of integrated luminosity. Thus much higher $\mchtwopm$ can indeed be probed even for $\mlspone$ beyond the reach of Run I. Moreover it is natural to expect that when the background is more accurately measured from the data the actual mass reach can be improved by optimizing the cuts.

In the absence of the above compression both lighter and heavier eweakinos can in principle contribute to the $3l$ signal. This is illustrated by the constraints derived for BP2 and BP3 (Table \ref{tab1}). It follows from these  examples  that for a fixed $M_2$  ($\mchonepm$) 
one can constrain the free parameter $\mchonepm$ ($\mchtwopm$ ) as is illustrated by BP2 (BP3). It is worth recalling that for decoupled heavier eweakinos there was no limit on the lighter eweakinos for $\mlspone \approx 100$ GeV as is the case 
in both the examples (see \cite{mc2}, Fig 5)\footnote{We confirm the validity of this result 
with the latest constraints \cite{atlas3l} which are somewhat stronger than the earlier ones used in \cite{mc2}.}. The main message of this analysis is that a large portion of the parameter space with non-decoupled $\chtwopm$, $\lspfour$ had in principle been  within the kinematical reach of the Run I experiments. It is, therefore, natural to seriously consider the possibility that they may show up even in the early phases of the  experiments at 13 TeV.  Especially if  a signal shows up,
both the lighter and heavier eweakinos  would demand serious attention in the race for revealing the underlying physics.     

Stronger new bounds are also obtained in  the Light Higgsino and light Left Slepton (LHLS) model
(Fig. 3b of \cite{mc2}). In this scenario only the left sleptons are assumed to be lighter than $\chonepm$ and $\lsptwo$. Following \cite{atlas3l} and \cite{mc2} their masses are chosen to be $\mslepl  = (\mchonepm + \mlspone)/2$. The limit 
$\mchonepm > $ 400.0 GeV corresponds  to BP4 (Table \ref{tab1}) with $M_2=1.5 \mu$ and $\mlspone$ = 170 GeV. For this $\mlspone$ and decoupled heavier eweakinos (i.e, $M_2$ having a significantly larger value) a much weaker bound $\mchonepm > $ 270.0 GeV was obtained (\cite{mc2}, Fig. 3b). 

In the Light Mixed and light Left Slepton (LMLS) model we have $M_2 \approx \mu$ and the LSP is a bino ($M_1 < \mu$) ( Fig. 4b in \cite{mc2}). The left slepton masses  are chosen as in the LHLS model. It follows from these  examples  that for a fixed $M_2$   
one can constrain the free parameter $\mchonepm$ (see BP5). 
 For the chosen LSP mass there is no limit on $\mchonepm$ for decoupled heavier eweakinos. 

\begin{table}[h]
\begin{center}
\begin{tabular}{|c|c|c|c|c|c|}
\hline
Parameters/ &BP1 &BP2 &BP3 &BP4 &BP5 \\ 
Masses &(LHHS) &(LHHS) &(LHHS) &(LHLS) &(LMLS)   \\
\cline{1-6}
$M_1 $   &191 &105 &105 &175 &296          \\
\hline
$\mu$   &$\simeq M_1$ &- &264 &- &$1.05 M_2$    \\
\hline
$M_2$ &-  &$1.5 \mu$  &- &$1.5 \mu$ &566           \\
\hline
$m_{\lspone}$ &152  &100  &100  &170  &290            \\
\hline
$m_{\chonepm}$ &178 &$>250$ &250 &$>400$ &$>540$     \\
\hline 
$m_{\chtwopm}$ &$>370$  &*  &$>415$ &* &*     \\
\hline

\end{tabular}
\end{center}
\caption{New (modified) limits on $\mchtwopm(\mchonepm)$ for fixed $\chonepm(M_2)$ in different models with non-decoupled $\chtwopm$ and $ \lspfour$. All masses and mass parameters are in GeV. '-' denotes that the corresponding mass parameters are treated as free
parameters and '*'  indicates  that the corresponding  $\mchtwopm$  is determined by $M_2$ and the lower limit on $\mchonepm$. The modified limits on $\mchonepm$ are stronger than the corresponding limits in the decoupled scenario.}
\label{tab1}
\end{table}

\begin{table}[h]
\begin{center}
\begin{tabular}{|c|c|c|c|c|c|c|c|c|}
\hline
Parameters/ & & & & & & &  &Total SM \\
Masses and &BP1 &BP2 &BP3 &BP4 &BP5 &BP6 &BP7 &Backgrounds  \\
Signals      &(LHHS) &(LHHS) &(LHHS) &(LHHS) &(LHHS) &(LHLS) &(LMLS) & \\
\cline{1-9}
$M_1 $   &191 &222 &132 &105 &104 &249 &321 &-        \\
\hline
$\mu$   &186 &268 &133 &270 &308  &300 &401 &-  \\
\hline
$M_2$ &350  &286  &486 &405 &462 &450 &382    & -     \\
\hline
$m_{\lspone}$ &151  &200  &100  &100  &100 &231 &305   &-     \\
\hline
$m_{\chonepm}$ &178 &234 &132 &260 &300  &291 &350  &- \\
\hline 
$m_{\chtwopm}$ &389  &351  &520 &447 &504 &491 &465  & - \\
  &(885) &(880) &(890) &(810) &(927) &(902) & & \\
\hline
3 leptons &73.8 &35.9 &107.7 &70.4 &56.4 &139.4 &58.2  &26.71  \\
  &(17.3) &(12.0) &(17.1) &(16.1) &(7.84) &(21.9) &(30.9) &   \\
\hline
$(S/\sqrt{B})_{3l}$ &14.3 &6.95 &20.8 &13.6 &10.9 &26.9 &11.3  &- \\
  &(3.35) &(2.32) &(3.31) &(3.12) &(1.52) &(4.24) &(5.98) & \\
\hline
4 leptons   &61.5 &52.5 &51.7 &16.4 &8.73 &19.6 &10.2    &0.835       \\
  &(0.69) &(1.20) &(-) &(0.62) &(0.36) &(2.05) &(-) & \\
\hline
SS3OS1 leptons  &29.9 &17.1 &14.5 &7.2 &3.36 &5.01 &1.57 &0.40    \\
  &(0.69) &(0.30) &(-) &(-) &(0.36) &(0.17) &(-) & \\
\hline
5 leptons &8.46  &8.29 &4.14 &6.1 &2.68 &4.14 &0.78   &0.60        \\
  &(-) &(0.60) &(-) &(-) &(-) &(-) &(-) &   \\
\hline

\end{tabular}
\end{center}

\caption{Number of $3l$, $4l$, $SS3OS1$, $5l$ events all with $\met$ corresponding to different BPs at LHC $13$TeV for integrated luminosity
of $100 fb^{-1}$ along with the total SM background in each case. The significance of the 3l signal is also shown for each BP. The contents of the brackets are numbers in the corresponding  decoupled scenario which are significantly smaller. All masses and mass parameters are in GeV.}
\label{tab2}
\end{table}

The above results encourage us to look into the multilepton +  $\met$ signatures in  models with non-decoupled heavier eweakinos at LHC 13 TeV experiments.
We begin with  the $ 4l + \met$ signal. It may be recalled that the ATLAS collaboration analysed this signal towards
 the end of Run I assuming decoupled heavier eakinos\cite{atlas4l} for a RPC simplified model assuming that the signal comes only from higgsino like $\lsptwo -\lspthree$ pair production. It was further assumed that they decay via any one of the following options: i)$R$-type selectrons or smuons, ii) staus or  iii) Z bosons with 100 \% BR. In contrast our broader framework considers all eweakino pair productions in several  generic MSSM models each represented by a BP displayed in Table \ref{tab2}. These BPs correspond to diverse compositions of the eweakinos, different 
mass hierarchies among the EW sparticles and realistic BRs of the relevant decay modes. All BPs are consistent with the new constraints 
derived in  this paper for non-decoupled $\chtwopm$ and $\lspfour$ ( Fig. 1 and Table \ref{tab1}). 

\begin{table}[h]
\begin{center}
\begin{tabular}{|c|c|c|c|c|c|c|c|c|c|c|c|}
\hline

Benchmark &$\sigma_{prod}$ &\multicolumn{3}{c|}{$\sigma_{eff}^{3l}$ in $fb$} &\multicolumn{3}{c|}{$\sigma_{eff}^{4l}$ in $fb$} &\multicolumn{2}{c|}{$\sigma_{eff}^{SS3OS1}$ in $fb$} &\multicolumn{2}{c|}{$\sigma_{eff}^{5l}$ in $fb$}  \\
\cline{3-5} \cline{6-8} \cline{9-10} \cline{11-12}
Points &in $pb$ &after &after &after &after &after &after &after &after &after &after \\
 & &$A1$ &$A2$ &$A3$ &$B1$ &$B2$ &$B3$ &$C1$ &$C2$ &$D1$ &$D2$ \\
\cline{1-12}

BP1 &769.1 &8.96 &7.54 &0.74 &1.42 &1.01 &0.62 &0.38 &0.30 &0.15 &0.08 \\
    &(691.6) & & & & & & & & & & \\
BP2 &553.0 &10.5 &8.09 &0.36 &1.68 &1.06 &0.51 &0.39 &0.17 &0.19 &0.07 \\
   &(300.7) & & & & & & & & & & \\
BP3 &2071.0 &7.08 &6.65 &1.08 &0.74 &0.62 &0.52 &0.16 &0.14 &0.06 &0.04 \\
  &(2060.0) & & & & & & & & & & \\
BP4 &380.8 &5.06 &2.87 &0.70 &0.45 &0.22 &0.16 &0.09 &0.07 &0.08 &0.06 \\
 &(309.1) & & & & & & & & & & \\
BP5 &223.7 &2.86 &1.67 &0.56 &0.28 &0.11 &0.09 &0.04 &0.03 &0.03 &0.026 \\
  &(182.3) & & & & & & & & & & \\
BP6 &217.9 &15.9 &14.6 &1.39 &0.51 &0.40 &0.20 &0.06 &0.05 &0.05 &0.04 \\
  &(170.9) & & & & & & & & & & \\ 
BP7 &156.9 &12.3 &11.1 &0.58 &0.30 &0.19 &0.10 &0.02 &0.015 &0.03 &0.0078 \\
  &(72.6) & & & & & & & & & & \\
\hline

\end{tabular}
\end{center}
\caption{The production cross sections of all eweakino pairs and the effective cross-section  after successive cuts of four types of signals for the BPs  defined in Table \ref{tab2}. The contents of the brackets are numbers in the corresponding  decoupled scenarios. }
\label{tab3}
\end{table}

\begin{table}[h]
\begin{center}
\begin{tabular}{|c|c|c|c|c|c|c|c|c|c|c|c|}
\hline

Background &$\sigma_{prod}$ &\multicolumn{3}{c|}{$\sigma_{eff}^{3l}$ in $fb$} &\multicolumn{3}{c|}{$\sigma_{eff}^{4l}$ in $fb$} &\multicolumn{2}{c|}{$\sigma_{eff}^{SS3OS1}$ in $fb$} &\multicolumn{2}{c|}{$\sigma_{eff}^{5l}$ in $fb$}  \\
\cline{3-5} \cline{6-8} \cline{9-10} \cline{11-12}
Processes &in $pb$ &after &after &after &after &after &after &after &after &after &after \\
 & &$A1$ &$A2$ &$A3$ &$B1$ &$B2$ &$B3$ &$C1$ &$C2$ &$D1$ &$D2$ \\
\cline{1-12}

$WZ$ &32.69 &168.3 &13.11 &0.18 &- &- &- &- &- &- &- \\
$ZZ$ &10.63 &16.5 &1.25 &0.007 &14.2 &0.081 &0 & &- & &- \\
$t \bar{t}Z$ &0.018 &1.95 &0.39 &0.015 &0.26 &0.039 &0.018 &0.006 &0.002 &0.002 &0.0007  \\
$WWZ$ &0.133 &1.33 &0.17 &0.013 &0.18 &0.012 &0.004 &- &- &- &-  \\
$WZZ$ &0.042 &0.54 &0.044 &0.005 &0.068 &0.0014 &0.0003 &0.007 &0.003 &0.013 &0.005 \\
$ZZZ$ &0.010 &0.05 &0.003 &0.0001 &0.04 &0.0003 &0.00005 &0.0004 &0.00003 &0.001 &0.0003  \\
$WWW$ &0.159 &0.79 &0.07 &0.059 &- &- &- &- &- & &- \\
\hline

\end{tabular}
\end{center}
\caption{ The production and effective cross-sections of different SM backgrounds for the four different 
signals. '-' denotes that the concerned  background process is not relevant for the  signal.}
\label{tab4}
\end{table}

An obvious physics background in this case  is  ZZ production. We have generated ZZ + 1 jet events
with MLM matching \cite{mlm}  using ALPGEN(v2.1) \cite{alpgen} which are then passed to PYTHIA for showering and jet formation using the anti-$k_t$ algorithm \cite{antikt}. 
We have simulated the signal  and all  SM backgrounds by selecting  events with

\noindent
B1) 4 isolated leptons consistent with the selection cuts mentioned above,\\
B2) Invariant mass of any pair of oppositely charged leptons of same flavour not in the window 81.2 $< m_{inv} <$ 101.2GeV, \\
and\\
B3) $\met >$ 80.0 GeV.\\ 

In Table II we have presented the relevant parameters defining each BP in rows 2-7. The number of 4l events N(4l) for 100 $\ifb$ of integrated luminosity subject to the above cuts for each BP and the total SM background are in row 9. For a better understanding of these numbers the total production cross section of all chargino neutralino pairs in each case  and the corresponding effective cross sections ($\sigma_{eff}^{4l}$) after the cuts B1) - B3) are given in columns 2 and 6-8 Table III. The total background cross section and the effective cross sections afer the cuts for different channels are in  Table IV. The total background is indeed tiny. If we require at least five signal events over a negligible background for a discovery, then  optimistic results are obtained for all BPs.   
On the other hand the number in parenthesis below each N(4l) stands for the corresponding number in the decoupled scenario. The numerical results in the non-decoupled (decoupled) models are obained for $ M_2 = 1.5 \mu$ ($ M_2 = 2 \mu$. It is clear that in a variety of
decoupled models the N(4l) is indeed negligible.

Two comments are now in order. 
For the $t \bar(t) Z$ a NLO corrected cross-section boosted by  a K-factor of 1.35 \cite{ttzNLO} yields about 5 background events. In order suppress it further we have used an additional cut. We reject events with at least one tagged $b$-jet following the criteria MV1 of \cite{btagg} and the effective cross-section in Table IV is reduced to 0.004 fb. The signal is hardly affected by this additional cut.
The irreducible backgrounds being negligible one has to look for the reducible backgrounds arising due to  jets faking leptons. Without a thorough detector simulation it is difficult to estimate this background. The analysis of
\cite{atlas4l}, however,  found this background to be negligible for the $4l +\met$ signal. It is, therefore, reasonable to assume that this background is negligible for all the signals with  four or more leptons considered in this paper. 

For comparison we also present in Table II the number of  3l + $\met$ events N(3l) obtained with the cuts 
A1) - A3) defined above and the total SM background for an integrated luminosity
of $100 fb^{-1}$ (row 8). The production cross section of all chargino-neutralino pairs, the effective cross sections after the cuts for both the signal the total background etc are also included in Tables III - IV following the same convention as in the $4l$ case. 
It readily follows from Table II the ratio $ N_{4l}/ N_{3l}$, which is free from several theoretical uncertainties, one can discriminate between many non-decoupled and decoupled models since the ratio is tiny in a wide variety of decoupled models. The same observable may also be useful for discriminating among the non-decoupled models. Similar relative rates involving other final states (see below) can also be used to facilitate this discrimination.  

The same methodology has been followed for generating the $SS3OS1+\met$ signal which is a subset of the 
$4l +\met$ events. However, this choice of the final state  significantly reduces the backgrounds involving multiple Z bosons or $t \bar{t} Z$. The main irreducible SM background in this  case are WZZ events where a lepton from any Z boson decay fails to pass the selection cuts. The selection cuts ($C_1$) and the cut $\met > 80$GeV ($C_2$) suppress this and other backgrounds listed in Table IV to negligible levels. The number of signal events for an integrated luminosity of 100 $\ifb$ corresponding to the above BPs are displayed in Table \ref{tab2}. The relevant information about the effective signal cross sections can be gleaned from the table \ref{tab3}. It may be noted that the relative rates of $4l$ and $SS3OS1$ events can  distinguish among different decoupled models.

The next entry in our list is the $5l + \met$ signal,  where $l$ stands for an $e$ or $\mu$ of any charge. The selection cuts (D1) and the requirement $\met > 80$ GeV (D2) cut suppress all the  backgrounds including the  potentially dangerous 
contribution  from WZZ events to a negligible level. 
We quote the number of signal events for the BPs studied  and the total background for  an integrated luminosity of 100 $\ifb$  
in Table \ref{tab2}.   

We now briefly comment on the signals in the LWLS model which yielded the strongest bounds on the lighter eweakinos (See Fig. 7a of \cite{atlas3l}). For $\mlspone \leq 250$GeV one obtains  $\mchonepm \geq 700$ GeV. In this case the heavier eweakinos are too massive to produce any observable signal before the LHC luminosity upgrade. However, if the lighter eweakino spectrum is to some extent compressed the above stringent bound on $\mchonepm$is relaxed. This is illustrated by the following parameter set:\\
$M_1 = 298.0 $ , $M_2 = 345.0$, $\mu = 518.0$, $\mlspone = 290.0$, $\mchonepm = 349.0$ and $\mchtwopm = 545.0$ (all in GeV).
In this scenario the number of  $4l$ events  and  $SS3OS1$ events are respectively 9.37  and 3.33 for 100 $\ifb$ of integrated luminosity with the above cuts.

The  potentially rich phenomenology of the heavier eweakinos calls  for further investigations in the light of the upcoming LHC data, the observed DM relic density of the universe and the $\gmin2$ anomaly. We have already checked that they may significantly  contribute to $\gmin2$. Further details will be provided elsewhere. \\
{ \bf Acknowledgments :} The research of AD was supported by the Indian National Science Academy, New Delhi.  NG thanks the Board of Research in Nuclear Sciences, Department of Atomic Energy, India for a research fellowship.


\end{document}